\documentclass[letter,modern,twocolumn,tighten]{aastex701}
\usepackage[english]{babel}
\usepackage[margin=0.6in]{geometry}
\usepackage{newtxtext,newtxmath}
\usepackage{comment}
\usepackage{multirow}
\usepackage{amsmath}
\usepackage{rotating}
\usepackage{longtable}
\usepackage[capitalize]{cleveref}

\begin{document}

\title{Search for ultra-high energy neutrons from Galactic sources with the Pierre Auger Observatory}

% created on 2025-10-10
\author[orcid=0000-0003-1218-1737]{A.~Abdul Halim}
\affiliation{University of Adelaide, Adelaide, S.A., Australia}
\email{adila.abdulhalim@adelaide.edu.au}

\author[orcid=0000-0002-9973-7314]{P.~Abreu}
\affiliation{Laborat\'orio de Instrumenta\c{c}\~ao e F\'\i{}sica Experimental de Part\'\i{}culas -- LIP and Instituto Superior T\'ecnico -- IST, Universidade de Lisboa -- UL, Lisboa, Portugal}
\email{abreu@lip.pt}

\author[orcid=0000-0001-8354-5388]{M.~Aglietta}
\affiliation{Osservatorio Astrofisico di Torino (INAF), Torino, Italy}
\affiliation{INFN, Sezione di Torino, Torino, Italy}
\email{aglietta@to.infn.it}

\author[orcid=0000-0002-9624-3044]{I.~Allekotte}
\affiliation{Centro At\'omico Bariloche and Instituto Balseiro (CNEA-UNCuyo-CONICET), San Carlos de Bariloche, Argentina}
\email{ingo@cab.cnea.gov.ar}

\author[orcid=0000-0001-6352-5339]{K.~Almeida Cheminant}
\affiliation{Nationaal Instituut voor Kernfysica en Hoge Energie Fysica (NIKHEF), Science Park, Amsterdam, The Netherlands}
\affiliation{IMAPP, Radboud University Nijmegen, Nijmegen, The Netherlands}
\email{kevin.cheminant@ru.nl}

\author[orcid=0000-0003-1233-2670]{A.~Almela}
\affiliation{Instituto de Tecnolog\'\i{}as en Detecci\'on y Astropart\'\i{}culas (CNEA, CONICET, UNSAM), Buenos Aires, Argentina}
\affiliation{Universidad Tecnol\'ogica Nacional -- Facultad Regional Buenos Aires, Buenos Aires, Argentina}
\email{alejandro.almela@iteda.cnea.gov.ar}

\author{R.~Aloisio}
\affiliation{Gran Sasso Science Institute, L'Aquila, Italy}
\affiliation{INFN Laboratori Nazionali del Gran Sasso, Assergi (L'Aquila), Italy}
\email{roberto.aloisio@gssi.it}

\author[orcid=0000-0002-2367-0803]{J.~Alvarez-Mu\~niz}
\affiliation{Instituto Galego de F\'\i{}sica de Altas Enerx\'\i{}as (IGFAE), Universidade de Santiago de Compostela, Santiago de Compostela, Spain}
\email{jaime.alvarez@usc.es}

\author[orcid=0000-0002-9942-1029]{A.~Ambrosone}
\affiliation{Gran Sasso Science Institute, L'Aquila, Italy}
\email{antonio.ambrosone@gssi.it}

\author[orcid=0000-0002-2059-6206]{J.~Ammerman Yebra}
\affiliation{Instituto Galego de F\'\i{}sica de Altas Enerx\'\i{}as (IGFAE), Universidade de Santiago de Compostela, Santiago de Compostela, Spain}
\email{juan.ammerman.yebra@usc.es}

% ...

\collaboration{all}{The Pierre Auger Collaboration}
\email{spokespersons@auger.org}

%\author[orcid=0000-0002-2245-5108]{L.~Caccianiga}
%\affiliation{INFN, Sezione di Milano, Milano, Italy}
%\affiliation{Universit\`a di Milano, Dipartimento di Fisica, Milano, Italy}
%\email{lorenzo.caccianiga@mi.infn.it}

\begin{abstract}
Deflections in the propagation of charged ultra-high-energy cosmic rays (UHECRs) caused by magnetic fields make the identification of their sources challenging.
On the other hand, the arrival directions at Earth of neutrons point directly to their origin.
The emission of UHECRs from a source is expected to be accompanied by the production of neutrons in its vicinity through interactions with ambient matter and radiation.
Since free neutrons travel a mean distance $d/\text{kpc}=9.2(E/\text{EeV})$ before decaying, a neutron flux in the EeV range could be detected on Earth from sources of UHECRs in our Galaxy.
Using cosmic-ray data from the Phase\,I of the Surface Detector of the Pierre Auger Observatory, we search for neutron fluxes from Galactic candidate sources.
We select more than 1000 objects of astrophysical interest, stacking them into target sets.
The targets all have declinations within the exposure of the Observatory, ranging from $-90^\circ$ up to $+45^\circ$ for energies above 1\,EeV (and up to $+20^\circ$ for energies down to 0.1\,EeV).
Given that a neutron air shower is indistinguishable from a proton one, there is a significant background due to cosmic rays.
A neutron flux from the direction of a candidate source would be identified by a celestial density of events that significantly exceeds the expected density of cosmic rays for that direction.
No significant excess is found at any tested target direction, and an upper limit on the neutron flux is calculated for each candidate source.
\end{abstract}

%\keywords{Galactic astroparticle physics, ultra-high-energy cosmic rays, ultra-high-energy neutrons, Pierre Auger Observatory}

\section{Introduction}
\label{sec:intro}

Neutral particles can be used to study sources of ultra-high-energy cosmic rays since they are not deflected by magnetic fields and their arrival directions on Earth point directly to their origin.
Neutral particles are expected to be produced in the vicinity of the sources of UHECRs.
In particular, ultra-high-energy (UHE) neutrons are expected to be produced near any source accelerating protons and nuclei due to nuclear collisions, pion photoproduction, and photodisintegration of UHE nuclei.
Although free neutrons are unstable, with a mean lifetime of ${\sim}878$\,s~\citep{navas2024review}, in the ultra-relativistic regime they travel a mean distance $d$ of around $d/\text{kpc}\approx9.2(E/\text{EeV})$ before decaying.
This range is comparable to typical Galactic scales, so a neutron flux from Galactic sources could be detected in the EeV range.

Among neutral particles, photons are easier to distinguish from the bulk of the nuclei cosmic rays, as the air showers they induce in the atmosphere have different characteristics.
Indeed, extensive studies have been carried out to search for photon-like events at the highest energies, with no detection so far (see \citet{PhysRevD.110.062005} and references therein).
On the other hand, neutrons induce air showers that are indistinguishable from those generated by protons.
However, neutron fluxes can be identified in an indirect way through an excess of events around the direction of their sources \footnote{Note that an EeV proton deflects about $\sim50^\circ$ while traveling 1\;kpc in a transverse field of $\mathcal{O}(\mu\text{G})$, so we do not expect significant excesses due to protons at these energies.}.
We search for such excesses, comparing the observed cosmic-ray (CR) density with the one expected from an isotropic distribution, as done in previous studies by the Pierre Auger Collaboration \citep{abreu2012search,aab2014targeted}, albeit with a different method. 

The dominant processes to create energetic neutrons and photons are pion-production interactions in which a UHE charged hadron interacts with ambient photons or nuclei.
Neutrons are produced in charge exchange interactions, while photons are the result of the decay of neutral pions (for a review on the neutron production, see e.g.~\citep{medina2001anisotropy,Bossa:2003fa,cavasinni2006supernova,anchordoqui2007neutrinos, Aartsen_2016}).
The likelihood of charge-exchange interactions that produce neutrons is of the same order of magnitude as that of producing neutral pions.
On average, neutral pions carry only a small fraction of the parent proton energy, whereas a neutron acquires most of it.
Neutrons can also be produced by photodisintegration of CR nuclei.
In this case, the photon emitted typically carries an energy about three orders of magnitude lower than the ejected neutron~\citep{anchordoqui2007photodisintegration}.
Therefore, we expect the hadronic production of neutrons to exceed that of photons at the same energy.
Moreover, muon-poor showers initiated by photons are typically reconstructed with energies lower than the actual energy of the primary photon because the reconstruction methods based on the measurements of the shower size are calibrated on the bulk of the measured showers, which are induced by nuclei~\citep{abreu2023photons}.
In contrast, neutron-induced showers are indistinguishable from those initiated by protons and are thus reconstructed at their true energy.
For these reasons, a source of UHECRs is more detectable by a flux of neutrons than by a flux of photons if the source is within the neutron-decay distance.

Previous searches carried out by the Pierre Auger Collaboration did not detect any excess ascribable to a neutron flux~\citep{abreu2012search,aab2014targeted}.
In this work, we present updated results of the targeted search for point sources of neutrons, extending the maximum zenith angle from $60^\circ$ to $80^\circ$ for events above 1\,EeV, which allows us to reach declinations up to $45^\circ$.
We also lower the energy threshold, investigating events with energies between 0.1 and 1\,EeV with zenith angles up to $55^\circ$, reaching declinations up to $+20^\circ$.
Additionally, in this work, we update the analysis method compared to previous studies.
We compare the celestial density of cosmic rays observed at the direction of the target with the density expected from an isotropic distribution.
However, instead of using a top-hat function to define an acceptance solid angle, as in previous analyses, we compute, and sum, the probability density of every event at the direction of a targeted source, accounting for the arrival direction of each event as well as an estimate of the event-by-event angular uncertainty.
The sum of all the event probability functions is the celestial cosmic ray density function (particles per unit solid angle).

\section{Datasets}
\label{sec:data}

The Pierre Auger Observatory is the largest facility in the world designed to study UHECRs.
It is located in Argentina, at a latitude of ${\sim}-35^\circ$.
The events used in this work were recorded by the Surface Detector (SD) array from 01 January 2004, up to 31 December 2022, the so-called Phase\,I of the Observatory.
The SD is an array of 1660 water-Cherenkov detectors (WCDs) arranged in a triangular grid, covering an area of approximately 3000\,km$^2$ with a duty cycle ${\sim}100\%$~\citep{aab2015observatory}.
While the spacing of 1600 WCDs is 1500\,m, 60 additional stations were deployed in an area of around 24\,km$^2$, with a reduced spacing of 750\,m in this sub-array.
Hereafter, the dataset recorded with this portion of the array will be referred to as the SD-750 dataset, while the one with 1500\,m spacing will be referred to as SD-1500.
For both datasets, the most stringent selection criteria are used, accepting events in which all six nearest neighboring grid positions of the station with the highest signal were occupied by stations that were operational \footnote{For events with zenith angle between $60^\circ$ and  $80^\circ$, where the stations surrounding the core are likely to have similar signals, the reference station is the one closest to the reconstructed shower core rather than the one with the highest signal.
This \textit{a posteriori} selection is introduced to ensure that the choice of the reference station is robust wrt statistical fluctuations of the signal}. 

In this analysis, we consider events recorded with the SD-1500 array with an energy above 1\,EeV and zenith angles up to $80^\circ$.
Events with zenith angles up to $60^\circ$ are reconstructed with a different algorithm ~\citep{aab2020reconstruction} than those in the zenith range from $60^\circ$ to $80^\circ$ ~\citep{augercollaboration2014inclined}.
In the following, we will refer to the former as \textit{vertical} and the latter as \textit{inclined} events.

The total exposure of the SD-1500 array, considering the full Phase\,I of the Observatory and after applying quality cuts, is 110,000\,km$^2$\,sr\,yr, yielding 2,654,574 events with $E \geq 1$\,EeV.
Following previous studies by the Pierre Auger Collaboration on the search for point sources of neutrons, we divide the dataset into three independent energy ranges: $1\,\text{EeV}\leq E<2\,\text{EeV}$ (2,009,321 events), $2\,\text{EeV}\leq E<3\,\text{EeV}$ (382,576 events), and $E\geq 3\,\text{EeV}$ (262,677 events).
The analysis is performed separately on these energy intervals, as well as on the cumulative range $E\geq 1\,\text{EeV}$.

The SD-750 dataset contains events recorded from 01 August 2008 to 31 December 2022.
The total exposure of the SD-750 array, after applying the quality cuts and including zenith angles up to $55^\circ$, is 408\,km$^2$\,sr\,yr, resulting in 1,455,168 events between 0.1\,EeV and 1\,EeV.
This dataset has been divided into three independent energy ranges: $0.1\,\text{EeV}\leq E<0.2\,\text{EeV}$ (1,069,076 events), $0.2\,\text{EeV}\leq E<0.3\,\text{EeV}$ (236,367 events), and $0.3\,\text{EeV}\leq E\leq 1\,\text{EeV}$ (149,725 events).
The analysis is performed in these three energy ranges plus the cumulative dataset with events with energies between 0.1 and 1\,EeV.

\section{Target sets}
\label{sec:Target}

We search for neutron emission from astrophysical candidate sources.
These sources are divided into 12 different target sets.
We selected candidate sources with declinations up to $+45^\circ$, resulting in a total of 1092 target directions investigated using the SD-1500 dataset.
For the analysis using the SD-750 dataset, we select sources within a measured distance of 1\,kpc, ensuring that the neutron flux can reach Earth within approximately one attenuation length across all energy ranges.
Given that the SD-750 dataset contains events with zenith angle up to $55^\circ$, only sources with declination less than $+20^\circ$ can be considered.
Therefore, of the 1092 sources selected for the SD-1500 analysis, 70 met the criteria and were included in the analysis with the SD-750 dataset.
If a candidate source is in more than one catalog, we consider it only in the most exclusive target set, i.e., the one with fewer objects.
Based on the angular resolution of the Observatory~\citep{aab2020reconstruction}, whenever multiple targets from the same set are found within less than $0.2^\circ$ we treat them as one single effective source. This happens in particular in the case of multiple sources in the same star cluster. Indeed, in almost all cases the quoted distance in the reference catalog was the same for the merged targets.
The complete list of the target sets used is:
\begin{itemize}
    \item Millisecond pulsars (msec PSRs)~\citep{manchester2005pulsar}: we consider 196 targets, of which 23 are considered in the analysis below 1\,EeV;
    \item Gamma-ray pulsars as observed by the Fermi Satellite ($\gamma$-ray PSRs)~\citep{abdollahi2020fermi}: we use an updated catalog version with respect to \citet{aab2014targeted}, including the new data available.
    We select 261 targets, of which 33 are considered in the analysis below 1\,EeV;
    \item Low- and High-mass X-ray binaries (LMXB and HMXB), from~\citet{LMXB} and ~\citet{HMXB}, respectively.\footnote{We obtained the flux in the X-band from the cross-referenced Swift 2SXPS \citep{2SXPS} counterparts.}
    In total, we consider 267 LMXBs and 139 HMXBs, and 6 and 2, respectively, for the analysis below 1\,EeV;
    \item Three classes of TeV gamma-ray emitters, selected from the TeVCat\,2.0 catalog~\citep{wakely2008tevcat}, choosing Galactic objects.\footnote{\url{http://tevcat2.uchicago.edu/}}
    The three classes are: Pulsar Wind Nebulae (TeV $\gamma$-ray - PWNe), other identified TeV gamma-ray sources (TeV $\gamma$-ray - other), and unidentified TeV gamma-ray sources (TeV $\gamma$-ray - UNID).
    These catalogs have been updated since \citet{aab2014targeted}, in which only H.E.S.S.\ data was used.
    In total, 31 TeV $\gamma$-ray - PWNe, 44 TeV $\gamma$-ray - other and 74 TeV $\gamma$-ray - UNID targets are tested.
    For the analysis below 1\,EeV, 3 PWN and 3 other identified sources are used;
    \item Microquasars, from S.~Chaty:\footnote{\url{www.aim.univ-paris7.fr/CHATY/Microquasars/microquasars.html}} of the 15 sources considered, none are tested in the energy interval below 1\,EeV;
    \item Magnetars, from~\citet{olausen2014magnetar,zelati2017magnetar}: this catalog has been updated since \citet{aab2014targeted}.
    We test 27 sources, none of which are considered at energies below 1\,EeV;
    \item Gamma PeVatrons, as labeled in the TeVCat\,2.0 catalog, including sources detected by the Large High Altitude Air Shower Observatory (LHAASO)~\citep{cao2024LHAASO}: it is the first time this catalog is considered in the search for sources of EeV neutrons.
    Of the 36 sources tested, 1 is considered for the energy interval below 1\,EeV;
    \item Galactic Center (GC);
    \item Crab pulsar~\citep{wakely2008tevcat}: this target was not considered in ~\citet{aab2014targeted} since events with zenith angles greater than $60^\circ$ were not included, leaving the Crab pulsar outside the field of view of that analysis.
\end{itemize}

%Based on the angular resolution of the Observatory~\citep{aab2020reconstruction}, whenever multiple targets from the same set are found within less than $0.2^\circ$ and at the same distance, as for multiple objects in the same cluster, we treat them as one single effective source.
The position of the effective source is taken to be that of the brightest of the individual objects, while its flux corresponds to the sum of their fluxes.

\section{Method}
\label{sec:method}

A neutron flux can be identified as an excess of events around its source direction.
To search for excesses, we compare the observed CR density to the expected density from an isotropic distribution of arrival directions, considering the exposure of the Observatory, following the method introduced in~\citet{franco2023neutron}.
The CR density is calculated as the sum of the probability densities of all events in the dataset.
This density is evaluated in each target direction.
The probability density of the $i$-th event evaluated at the direction of the $j$-th target is given by
\begin{equation}
  w_{ij} =
    \frac{1}{2\pi\sigma_i^2} \exp\left(-\frac{\xi_{ij}^2}{2\sigma_i^2}\right),
\label{density}
\end{equation}
where $\xi_{ij}$ is the angular distance between the arrival direction of $i$-th event and the direction of the $j$-th target, and $\sigma_i$ is the angular uncertainty of the event.
Although the reconstruction provides an estimate of the angular uncertainty for each event, these estimates can exhibit fluctuations.
For this reason, the angular uncertainty is parameterized based on data in bins of zenith angle and multiplicity (the number of WCDs triggered during an event).
For the SD-1500 dataset, we use different parametrizations for vertical and inclined events.
For the events considered in this work, the resulting values of $\sigma_i$ are within the range $0.2^\circ$ to $1.4^\circ$.
The observed CR density at the position of the $j$-th target is then the sum of the probability densities over the $N$ events in the dataset, $\rho_j^\text{obs}=\sum_{i=1}^N w_{ij}$.

The signal expected from an isotropic distribution is estimated using simulated datasets based on a scrambling technique~\citep{cassiday1990sources}.
An event is simulated by randomly sampling two events from the dataset and extracting the arrival time from one of them and the zenith angle and the angular uncertainty from the other.
By sampling the zenith angle and the angular uncertainty from the same event, we ensure that the angular uncertainty distribution of the simulated datasets follows the observed one.
An azimuth angle is randomly sampled from a uniform distribution. Each scrambled dataset has the same number of events as the observed one and is treated as the observed one, evaluating the weight for each simulated event using \cref{density}, and then summing over all events to obtain the simulated CR density $\rho_j^\text{sim}$.
This technique accounts for the exposure of the Observatory, providing a reliable estimate of the background contribution.
We simulate 10,000 scrambled datasets.
The $p$-value associated to the $j$-th target is then defined as the fraction of these simulated datasets in which $\rho_j^\text{sim}$ was equal to or greater than $\rho_j^\text{obs}$.
When reporting the smallest $p$-value of a target set, one has to account for the multiple trials within the same target set.
We hence penalize the $p$-value using $p_j^\ast = 1 - (1 - p_j)^M$, where $M$ is the number of targets in the set.
The penalized $p$-value represents the probability of obtaining, by chance, a $p$-value less than or equal to $p_j$, given a set of $M$ $p$-values sampled from a uniform distribution between 0 and 1. 

A neutron flux upper limit is derived for each target as follows.
From the 10,000 simulated datasets, we have 10,000 simulated values for the density at the target, assuming that there is no flux of neutrons.
We can study how the density would increase with a neutron signal by adding artificially created neutron events from the target direction.
For each artificial event, a parametrized angular uncertainty is sampled from the dataset.
A two-dimensional Gaussian distribution of that uncertainty is used to sample an angular offset between the target and the arrival direction, and also to give the Gaussian density at the target direction based on that offset as in \cref{density}.
The resulting density from the artificial event is added to $\rho^\text{sim}$, and this is done separately for each of the 10,000 simulated densities $\rho^\text{sim}$.
The procedure is then repeated so that each $\rho^\text{sim}$ includes the contribution from 2 artificial events, then 3, and higher numbers.
Let $f_n$ denote the fraction of simulations with $n$ added artificial events for which the density is less than the observed density $\rho_j^\text{obs}$.
Note that $1-f_0$ is the $p$-value that was obtained with no artificial events added to simulations.
The fraction $f_n$ decreases as $n$ increases, and the procedure of adding artificial events terminates when 
\begin{equation}
  f_n  \leq (1 - \text{CL}) f_0.
\end{equation}
Here CL is the confidence level (e.g.\ $\text{CL} = 0.95$ for the 95\% confidence upper limit to be derived here), and $n$ is then the upper limit on the number of neutron events.
This follows Zech's definition of an upper limit $s_\text{UL}$ given by
\begin{equation}
  P(\leq k | b + s_\text{UL}) = (1 - \text{CL}) \, P({\leq}k | b),
\end{equation}
where $P({\leq}k | b)$ is the probability of obtaining a count $k$ or less given a mean background $b$.
The upper limit on the neutron flux is the upper limit on the number of neutrons divided by the directional exposure.
The directional exposure is obtained by dividing the expected density by the CR intensity.
The expected density is the average obtained from 10,000 scrambled datasets.
The CR intensity is obtained by integrating the spectral shape described in ~\citep{aab2020spectrum}.
From the neutron particle flux, we also compute the upper limit on the energy flux, assuming an $E^{-2}$ spectrum as in previous works~\citep{abreu2012search, aab2014targeted}.

In addition to analyzing the target directions individually, we perform a stacked analysis, as done in~\citet{aab2014targeted}, by calculating the combined significance of each class of candidate sources with an unweighted and a weighted combination of the individual $p$-values of the targets.
A class containing multiple targets with small $p$-values can have a combined $p$-value that is small compared to any of the individual $p$-values.
We use the Fisher formula~\citep{fisher1925statistical} to calculate the combined significance.
The chance probability of a product ($\Pi$) of $M$ $p$-values sampled from a uniform distribution to be lower than or equal to the product ($\Pi_\text{obs}$) of the $M$ actual $p$-values $p_j$ is
\begin{align}
\begin{split}
  \mathcal{P}(\Pi\leq\Pi_\text{obs}) &=
    \Pi_\text{obs}\sum_{j=0}^{M-1}\frac{(-\ln\Pi_\text{obs})^j}{j!} =
\\
    &= 1 - \text{Poisson}({\geq}M;-\ln\Pi_\text{obs}),
\label{fisher}
\end{split}
\end{align}
where $\text{Poisson}({\geq}k;\mu)$ is the Poisson probability of finding $k$ or more occurrences when the expected number is $\mu$.
We also evaluate the combined $p$-value including statistical weights.
The weight of a target is proportional to its directional exposure, to its measured electromagnetic flux, and to the attenuation factor due to neutron decay for its distance.
The statistical weights are proportional to these factors, and normalized in a way that the sum of all the weights in a target set is equal to 1.
We follow the same procedure to evaluate the weighted combined $p$-value as in~\citet{aab2014targeted}.
Since we do not have flux and distance information for some of the candidate sources, we calculate the weighted combined $p$-value using only those with complete information.\footnote{In the case of the Gamma PeVatrons and the unidentified TeV gamma-ray sources, the distance was not considered for the weight as that information was not available for the majority of the candidate sources in the target set.}
The unweighted combined $p$-value is evaluated for all candidate sources in each set.

\section{Results}
\label{sec:results}

We present the results of the combined analysis in \cref{tab:combined_results_SD1500,tab:combined_results_SD750}, for the SD-1500 and SD-750, respectively.
We report, separately for the unweighted and weighted combined analyses, the number of sources used in each target set as well as the combined $p$-value for different energy ranges.
The results for the most significant target in each set are presented in \cref{tab:result_vert+inc,tab:result_infill} for the SD-1500 and SD-750 datasets, respectively.
For each target reported, we present its direction in equatorial coordinates, the observed and expected CR density at its direction, the upper limit on the flux and the energy flux, the $p$-value, and the penalized $p$-value considering the total number of candidate sources in the target set.
None of the penalized $p$-values reached the $3\sigma$ threshold, indicating the absence of statistically significant evidence of a neutron flux coming from the tested directions across all energy ranges considered.  

\begin{table*}
    \setlength{\tabcolsep}{10pt}
    \centering
    \caption{Results for the combined analysis for different energy ranges using the SD-1500 dataset. The weighted analysis takes into account the distance and relative intensity of each source and is applied only to the targets where the relevant information is available.}
    \begin{tabular}{l|rrrrr|rrrrr}
		
%	\multirow{2}{*}{Class} & \multirow{2}{*}{No.} & \multicolumn{4}{c|}{Unweighted combined $p$-value} & No. & \multicolumn{4}{c}{Weighted combined $p$-value} \\ \cline{3-6} \cline{8-11}
%	&     &  $\geq$ 1\,EeV & $1-2$\,EeV & $2-3$\,EeV & $\geq$ 3\,EeV &  &  $\geq$ 1\,EeV & $1-2$\,EeV & $2-3$\,EeV & $\geq$ 3\,EeV \\
	\multirow{3}{*}{Class} & \multirow{3}{*}{No.} & \multicolumn{4}{c|}{Unweighted combined $p$-value} &  & \multicolumn{4}{c}{Weighted combined $p$-value} \\ 
    \cline{2-6} \cline{7-11}
	&     &  \multicolumn{4}{c|}{Energy range (EeV)} & No.  & \multicolumn{4}{c}{Energy range (EeV)}\\
 %    \cline{3-6} \cline{8-11}
	&     &  $\geq$ 1 & $1-2$ & $2-3$ & $\geq$ 3 &  &  $\geq$ 1 & $1-2$ & $2-3$ & $\geq$ 3 \\
	\hline
	\hline
		
	msec PSRs & 196 & 0.22 & 0.33 & 0.13 & 0.57  & 127 & 0.64 & 0.86 & 0.011 & 0.80 \\
$\gamma$-ray PSRs & 261 & 0.49 & 0.22 & 0.63 & 0.95  & 154 & 0.078 & 0.034 & 0.46 & 0.59 \\
LMXB & 267 & 0.79 & 0.93 & 0.0078 & 0.86  & 104 & 0.32 & 0.22 & 0.23 & 0.84 \\
HMXB & 139 & 0.78 & 0.77 & 0.15 & 0.91  & 79 & 0.51 & 0.57 & 0.90 & 0.11 \\
TeV $\gamma$-ray - PWNe & 31 & 0.86 & 0.85 & 0.38 & 0.73  & 23 & 0.027 & 0.0086 & 0.045 & 0.83 \\
TeV $\gamma$-ray - other & 44 & 0.59 & 0.85 & 0.11 & 0.40  & 19 & 0.16 & 0.49 & 0.26 & 0.094 \\
TeV $\gamma$-ray - UNID & 74 & 0.90 & 0.90 & 0.80 & 0.54  & 16 & 0.17 & 0.42 & 0.16 & 0.21 \\
Microquasars & 15 & 0.23 & 0.33 & 0.34 & 0.56  & 15 & 0.88 & 0.87 & 0.71 & 0.59 \\
Magnetars & 27 & 1.0 & 1.0 & 0.80 & 0.62  & 20 & 0.97 & 0.86 & 0.72 & 0.73 \\
$\gamma$-PeVatrons & 36 & 0.11 & 0.15 & 0.45 & 0.33  & 18 & 0.20 & 0.50 & 0.15 & 0.24 \\
Crab & 1 & 0.63 & 0.45 & 0.27 & 0.94  & $\cdots$ & $\cdots$ & $\cdots$ & $\cdots$ & $\cdots$ \\
Gal.\ Center & 1 & 0.88 & 0.84 & 0.61 & 0.69  & $\cdots$ & $\cdots$ & $\cdots$ & $\cdots$ & $\cdots$ \\

 \end{tabular}
 \label{tab:combined_results_SD1500}
\end{table*}

\begin{table*}
    \setlength{\tabcolsep}{8pt}
	\centering
	\caption{Results for the combined analysis for different energy ranges using the SD-750 dataset. The weighted analysis takes into account the distance and relative intensity of each source and is applied only to the targets where the relevant information is available.}
	\begin{tabular}{l|rrrrr|rrrrr}
%		\multirow{2}{*}{Class} & \multirow{2}{*}{No.} & \multicolumn{4}{c|	}{Unweighted combined $p$-value} & No. & \multicolumn{4}{c}{Weighted combined $p$-value} \\ \cline{3-6} \cline{8-11}
%		 &     &  $\Delta E_{0.1 - 1}$ & $\Delta E_{0.1 - 0.2}$ & $\Delta E_{0.2- 0.3}$ &  $\Delta E_{0.3 - 1}$ & & $\Delta E_{0.1 - 1}$ & $\Delta E_{0.1 - 0.2}$ & $\Delta E_{0.2 - 0.3}$ & $\Delta E_{0.3 - 1}$ \\
%\multirow{2}{*}{Class} & \multirow{2}{*}{No.} & \multicolumn{4}{c|}{Unweighted combined $p$-value} & No. & \multicolumn{4}{c}{Weighted combined $p$-value} \\ \cline{3-6} \cline{8-11}	&     &  $0.1-1$\,EeV & $0.1-0.2$\,EeV & $0.2-0.3$\,EeV & $0.3-1$\,EeV &  &  $0.1-1$\,EeV & $0.1-0.2$\,EeV & $0.2-0.3$\,EeV & $0.3-1$\,EeV \\
	\multirow{3}{*}{Class} & \multirow{3}{*}{No.} & \multicolumn{4}{c|}{Unweighted combined $p$-value} &  & \multicolumn{4}{c}{Weighted combined $p$-value} \\ 
    \cline{2-6} \cline{7-11}
	&     &  \multicolumn{4}{c|}{Energy range (EeV)} &  No.  & \multicolumn{4}{c}{Energy range (EeV)}\\
 %    \cline{3-6} \cline{8-11}
	&     &  $\geq$ 0.1 & $0.1-0.2$ & $0.2-0.3$ & $\geq$ 0.3 &  &  $\geq$ 0.1 & $0.1-0.2$ & $0.2-0.3$ & $\geq$ 0.3 \\
	\hline
	\hline

msec PSRs & 23 & 0.53 & 0.30 & 0.54 & 0.66  & 20 & 0.60 & 0.59 & 0.89 & 0.083 \\
$\gamma$-ray PSRs & 33 & 0.47 & 0.69 & 0.042 & 0.69  & 33 & 0.92 & 0.88 & 0.84 & 0.51 \\
LMXB & 6 & 0.73 & 0.81 & 0.83 & 0.45  & 5 & 0.92 & 0.89 & 0.94 & 0.31 \\
HMXB & 2 & 0.21 & 0.25 & 0.33 & 0.084  & 2 & 0.48 & 0.11 & 0.56 & 0.98 \\
TeV $\gamma$-ray - PWNe & 2 & 0.66 & 0.54 & 0.27 & 0.36  & 2 & 0.79 & 0.83 & 0.88 & 0.13 \\
TeV  $\gamma$-ray - other & 3 & 0.62 & 0.58 & 0.39 & 0.81  & 2 & 0.43 & 0.42 & 0.17 & 0.86 \\
$\gamma$-PeVatrons & 1 & 0.21 & 0.23 & 0.069 & 0.68  & $\cdots$ & $\cdots$ & $\cdots$ & $\cdots$ & $\cdots$ \\

 \end{tabular}
 \label{tab:combined_results_SD750}
\end{table*}

\begin{table*}
    \setlength{\tabcolsep}{8pt}
    \centering
    \caption{Results for the most significant target in each target set obtained using the SD-1500 dataset for the energy range above 1\,EeV. The upper limits are computed at 95\% confidence level. The penalization accounts for the number of targets tested in each target set.}
    \begin{tabular}{lrrrrrrrr}
    \multirow{2}{*}{Class} & R.A.  & Dec. & $\rho^{\mathrm{obs}}$ & $\rho^{\mathrm{exp}}$ & Flux UL & E-flux UL & \multirow{2}{*}{$p$-value} & $p$-value   \\
    & (deg) &  (deg) & (deg$^{-2}$) & (deg$^{-2}$) & (km$^{-2}$\,yr$^{-1}$) & (eV\,cm$^{-2}$\,s$^{-1}$) & & (penalized) \\
    \hline\hline
msec PSRs & 277.12 & 6.42 & 72.64 & 57.49 & 0.024 & 0.17 & 0.00070 & 0.13 \\
$\gamma$-ray PSRs & 296.63 & $-54.05$ & 144.60 & 124.22 & 0.014 & 0.10 & 0.00020 & 0.051 \\
LMXB & 117.14 & $-67.75$ & 160.57 & 140.62 & 0.013 & 0.095 & 0.0027 & 0.51 \\
HMXB & 212.01 & $-61.98$ & 147.51 & 131.02 & 0.012 & 0.086 & 0.0051 & 0.51 \\
TeV $\gamma$-ray - PWNe & 128.75 & $-45.60$ & 127.63 & 117.18 & 0.0096 & 0.070 & 0.025 & 0.55 \\
TeV $\gamma$-ray - other & 98.25 & 5.79 & 66.70 & 58.41 & 0.016 & 0.12 & 0.037 & 0.81 \\
TeV $\gamma$-ray - UNID & 305.02 & 40.76 & 13.35 & 7.25 & 0.090 & 0.65 & 0.020 & 0.78 \\
Microquasars & 308.11 & 40.96 & 13.84 & 7.01 & 0.099 & 0.73 & 0.011 & 0.15 \\
Magnetars & 275.57 & $-16.07$ & 96.35 & 91.29 & 0.0085 & 0.062 & 0.14 & 0.98 \\
$\gamma$-PeVatrons & 292.25 & 17.75 & 45.92 & 36.63 & 0.027 & 0.20 & 0.023 & 0.56 \\
Crab & 83.63 & 22.01 & 27.95 & 29.44 & 0.014 & 0.10 & 0.63 & 0.63 \\
Gal.\ Center & 266.42 & $-29.01$ & 98.71 & 104.25 & 0.0036 & 0.027 & 0.88 & 0.88 \\
    
    \end{tabular}
    \label{tab:result_vert+inc}
\end{table*}

\begin{table*}
    \centering
    \setlength{\tabcolsep}{8pt}
    \caption{Results for the most significant target in each target set obtained using the SD-750 dataset for the energy range between 0.1 and 1\,EeV. The upper limits are computed at 95\% confidence level. The penalization accounts for the number of targets tested in each target set.}
    \begin{tabular}{lrrrrrrrr}
    \multirow{2}{*}{Class} & R.A. & Dec. & $\rho^{\mathrm{obs}}$ & $\rho^{\mathrm{exp}}$ & Flux UL & E-flux UL & \multirow{2}{*}{$p$-value} & $p$-value   \\
    & (deg) & (deg) & (deg$^{-2}$) & (deg$^{-2}$) & (km$^{-2}$\,yr$^{-1}$) & (eV\,cm$^{-2}$\,s$^{-1}$) & & (penalized) \\
    \hline\hline
msec PSRs & 358.96 & 0.85 & 40.39 & 34.41 & 2.0 & 15.0 & 0.026 & 0.46 \\
$\gamma$-ray PSRs & 284.58 & $-22.28$ & 74.45 & 65.62 & 1.5 & 11.0 & 0.015 & 0.39 \\
LMXB & 292.71 & 5.52 & 27.56 & 25.39 & 1.7 & 12.0 & 0.20 & 0.73 \\
HMXB & 190.71 & $-63.06$ & 78.60 & 73.30 & 1.1 & 7.9 & 0.11 & 0.21 \\
TeV  $\gamma$-ray  PWNe & 98.12 & 17.37 & 4.47 & 4.20 & 3.8 & 28.0 & 0.37 & 0.61 \\
TeV $\gamma$-ray - other & 258.39 & $-39.76$ & 80.08 & 76.92 & 0.86 & 6.3 & 0.22 & 0.53 \\
$\gamma$-PeVatrons & 98.48 & 17.77 & 4.44 & 3.65 & 4.8 & 35.0 & 0.21 & 0.21 \\
    
    \end{tabular}
    \label{tab:result_infill}
\end{table*}

\begin{figure}
  \centering
  \includegraphics[width=\linewidth]{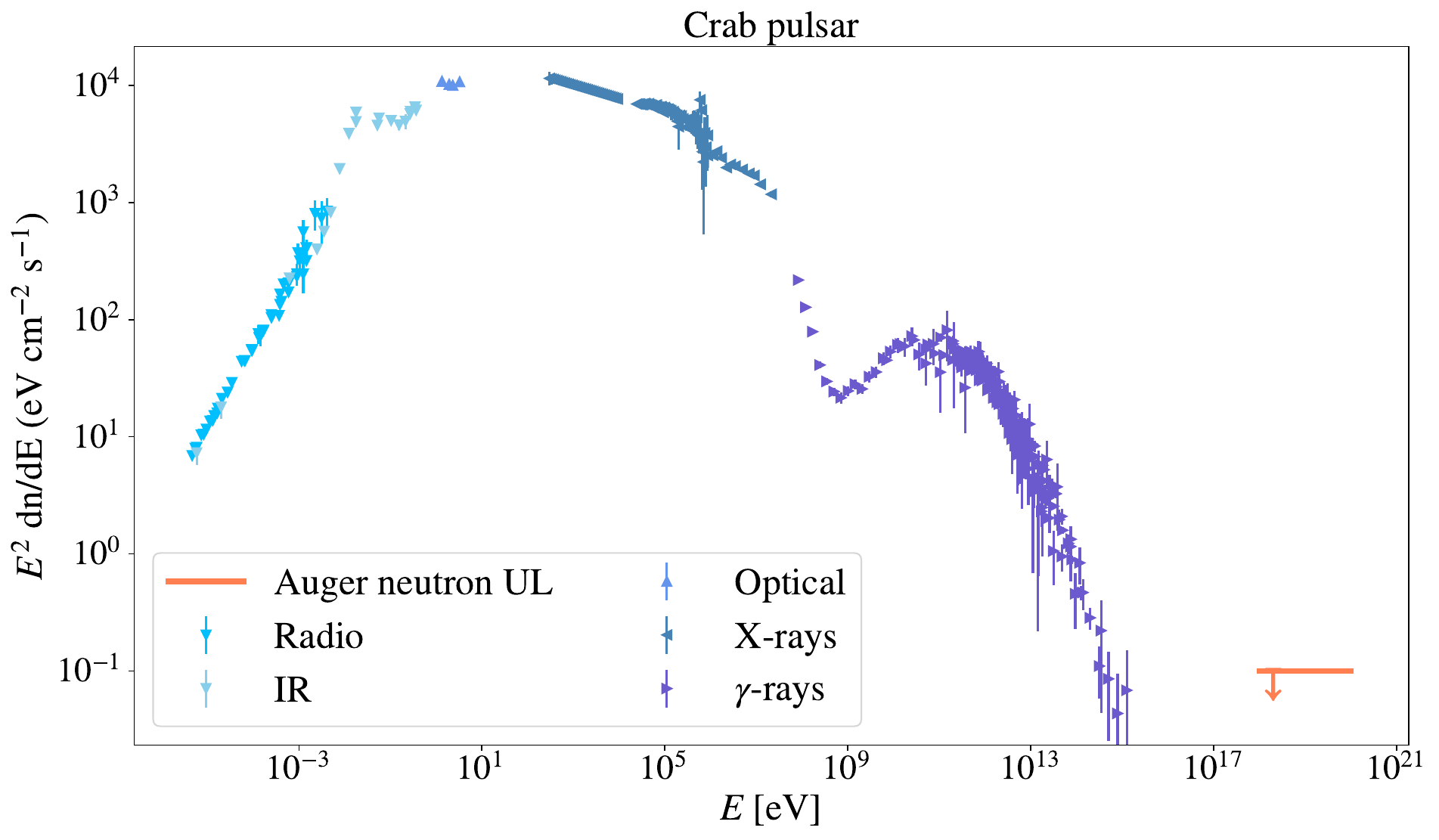}
  \\[3mm]
  \includegraphics[width=\linewidth]{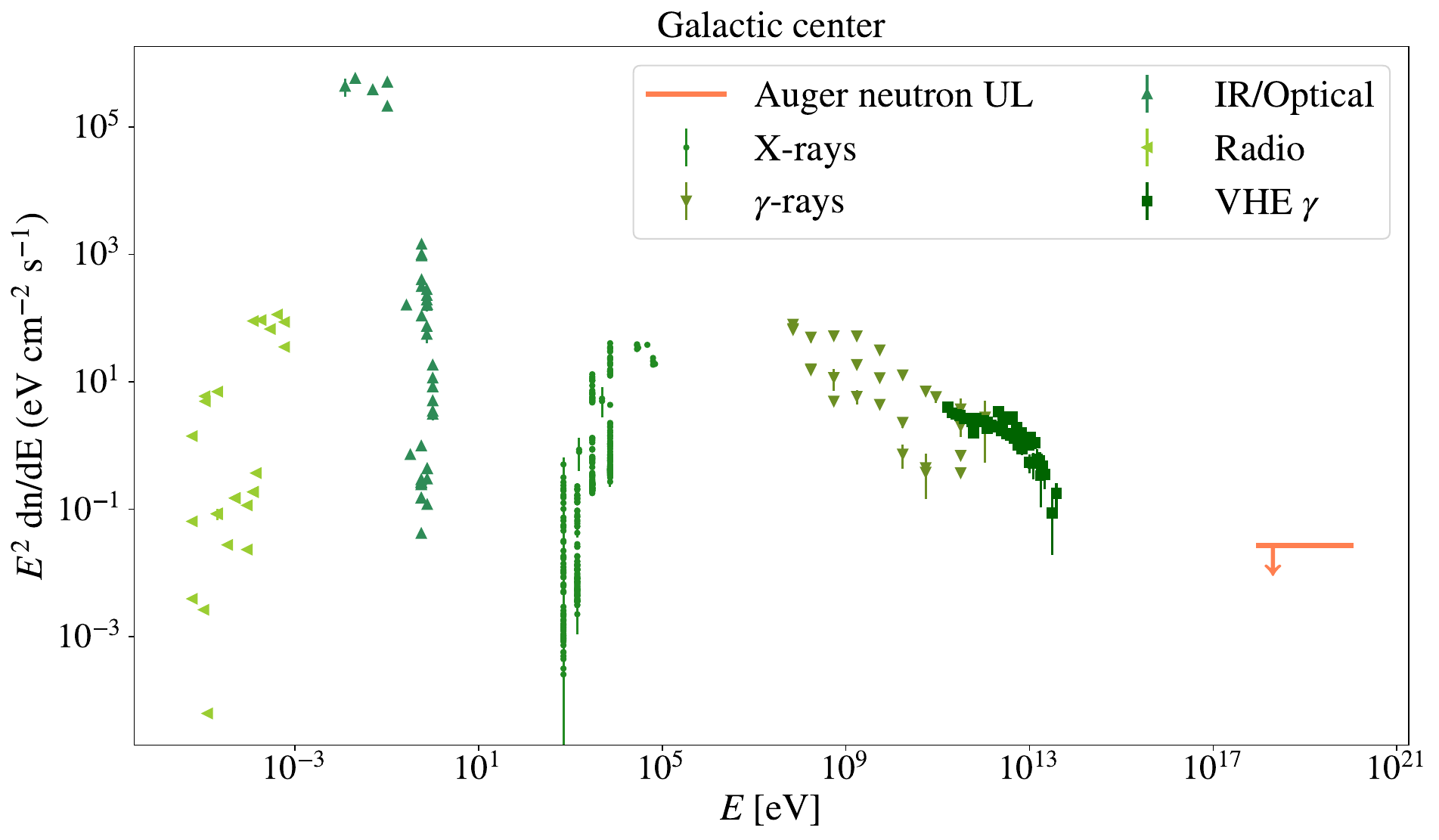}
  \caption{Spectral energy distributions (SEDs) for the Crab pulsar (top) and the Galactic center (bottom).
  For the Crab pulsar, the electromagnetic SED is from ~\citet{Pheno_Crab} with data available in ~\citet{Pheno_Crab_Data}.
  For the Galactic center, the majority of data is obtained through the tool SED Builder (\url{https://tools.ssdc.asi.it/SED}) with data from ~\citet{GC_Gamma,GC_IR,GC_Planck,GC_Radio,GC_VLA,GC_VizierJ} with the manual addition of VHE-$\gamma$ data from MAGIC \citep{GC_Magic}, Veritas \citep{GC_Veritas}, and H.E.S.S.\ \citep{GC_HESS}.}
  \label{fig:sed}
\end{figure}

The lack of statistically significant $p$-values has allowed us to place significant upper limits on the neutron emission from the targets, which can be compared to the observed electromagnetic emission.
In \cref{fig:sed}, we show the upper limits we obtain compared to the electromagnetic spectral energy distributions (SEDs) for the Galactic Center and the Crab pulsar.
In the supplementary material of this paper, machine-readable tables including the full results for all the tested targets are available.

\section{Conclusions}
\label{sec:conclusions}

In this work, we performed an updated search for neutron fluxes from more than 1000 Galactic sources.
We have not detected any statistically significant overdensity that would be associated with a neutron flux from any of the tested directions in the considered energy ranges.

We established upper limits of the neutron flux for all tested directions, considering a total exposure larger by a factor ${\sim}3$ compared to ~\citet{aab2014targeted}.
The sky coverage of the analysis was also extended, up to declination $+45^\circ$.
In addition, we reduced the energy threshold to 0.1\,EeV for sources located within the neutron decay length and declinations up to $+20^\circ$.
The upper limits presented in this work are, on average, improved by a factor of ${\sim}2$ compared to those reported in ~\citet{aab2014targeted} for the same targets.
These results are the most stringent direct constraints on hadronic acceleration in Galactic sources above 100\,PeV and can be used to constrain astrophysical models of production of UHECRs.

The aim of this work was to search for steady sources of neutrons in our Galaxy.
However, many of the targets tested in this analysis show variability in their electromagnetic emission.
Our time-averaged upper limits do not constrain a burst whose integrated flux is small compared to the total background for that target.
A search for flares or variability that matches what is observed through electromagnetic fluxes is forthcoming.
Additionally, we plan to perform an updated blind search over the whole sky, as presented in \citep{abreu2012search}, and a dedicated search along the Galactic Plane.

The Pierre Auger Observatory has recently entered its so-called Phase\,II, following the \emph{AugerPrime} upgrade~\citep{castellina2019augerprime}.
The enhanced observatory will improve our ability to distinguish showers produced by heavy nuclei.
Removing them will reduce the background, enhancing the sensitivity of future neutron searches.  

\clearpage

\onecolumngrid
\begin{acknowledgments}
% created on 2025-09-05
%\section*{Acknowledgments}

\begin{sloppypar}
The successful installation, commissioning, and operation of the Pierre
Auger Observatory would not have been possible without the strong
commitment and effort from the technical and administrative staff in
Malarg\"ue. We are very grateful to the following agencies and
organizations for financial support:
\end{sloppypar}

\begin{sloppypar}
Argentina -- Comisi\'on Nacional de Energ\'\i{}a At\'omica; Agencia Nacional de
Promoci\'on Cient\'\i{}fica y Tecnol\'ogica (ANPCyT); Consejo Nacional de
Investigaciones Cient\'\i{}ficas y T\'ecnicas (CONICET); Gobierno de la
Provincia de Mendoza; Municipalidad de Malarg\"ue; NDM Holdings and Valle
Las Le\~nas; in gratitude for their continuing cooperation over land
access; Australia -- the Australian Research Council; Belgium -- Fonds
de la Recherche Scientifique (FNRS); Research Foundation Flanders (FWO),
Marie Curie Action of the European Union Grant No.~101107047; Brazil --
Conselho Nacional de Desenvolvimento Cient\'\i{}fico e Tecnol\'ogico (CNPq);
Financiadora de Estudos e Projetos (FINEP); Funda\c{c}\~ao de Amparo \`a
Pesquisa do Estado de Rio de Janeiro (FAPERJ); S\~ao Paulo Research
Foundation (FAPESP) Grants No.~2019/10151-2, No.~2010/07359-6 and
No.~1999/05404-3; Minist\'erio da Ci\^encia, Tecnologia, Inova\c{c}\~oes e
Comunica\c{c}\~oes (MCTIC); Czech Republic -- GACR 24-13049S, CAS LQ100102401,
MEYS LM2023032, CZ.02.1.01/0.0/0.0/16{\textunderscore}013/0001402,
CZ.02.1.01/0.0/0.0/18{\textunderscore}046/0016010 and
CZ.02.1.01/0.0/0.0/17{\textunderscore}049/0008422 and CZ.02.01.01/00/22{\textunderscore}008/0004632;
France -- Centre de Calcul IN2P3/CNRS; Centre National de la Recherche
Scientifique (CNRS); Conseil R\'egional Ile-de-France; D\'epartement
Physique Nucl\'eaire et Corpusculaire (PNC-IN2P3/CNRS); D\'epartement
Sciences de l'Univers (SDU-INSU/CNRS); Institut Lagrange de Paris (ILP)
Grant No.~LABEX ANR-10-LABX-63 within the Investissements d'Avenir
Programme Grant No.~ANR-11-IDEX-0004-02; Germany -- Bundesministerium
f\"ur Bildung und Forschung (BMBF); Deutsche Forschungsgemeinschaft (DFG);
Finanzministerium Baden-W\"urttemberg; Helmholtz Alliance for
Astroparticle Physics (HAP); Helmholtz-Gemeinschaft Deutscher
Forschungszentren (HGF); Ministerium f\"ur Kultur und Wissenschaft des
Landes Nordrhein-Westfalen; Ministerium f\"ur Wissenschaft, Forschung und
Kunst des Landes Baden-W\"urttemberg; Italy -- Istituto Nazionale di
Fisica Nucleare (INFN); Istituto Nazionale di Astrofisica (INAF);
Ministero dell'Universit\`a e della Ricerca (MUR); CETEMPS Center of
Excellence; Ministero degli Affari Esteri (MAE), ICSC Centro Nazionale
di Ricerca in High Performance Computing, Big Data and Quantum
Computing, funded by European Union NextGenerationEU, reference code
CN{\textunderscore}00000013; M\'exico -- Consejo Nacional de Ciencia y Tecnolog\'\i{}a
(CONACYT) No.~167733; Universidad Nacional Aut\'onoma de M\'exico (UNAM);
PAPIIT DGAPA-UNAM; The Netherlands -- Ministry of Education, Culture and
Science; Netherlands Organisation for Scientific Research (NWO); Dutch
national e-infrastructure with the support of SURF Cooperative; Poland
-- Ministry of Education and Science, grants No.~DIR/WK/2018/11 and
2022/WK/12; National Science Centre, grants No.~2016/22/M/ST9/00198,
2016/23/B/ST9/01635, 2020/39/B/ST9/01398, and 2022/45/B/ST9/02163;
Portugal -- Portuguese national funds and FEDER funds within Programa
Operacional Factores de Competitividade through Funda\c{c}\~ao para a Ci\^encia
e a Tecnologia (COMPETE); Romania -- Ministry of Research, Innovation
and Digitization, CNCS-UEFISCDI, contract no.~30N/2023 under Romanian
National Core Program LAPLAS VII, grant no.~PN 23 21 01 02 and project
number PN-III-P1-1.1-TE-2021-0924/TE57/2022, within PNCDI III; Slovenia
-- Slovenian Research Agency, grants P1-0031, P1-0385, I0-0033, N1-0111;
Spain -- Ministerio de Ciencia e Innovaci\'on/Agencia Estatal de
Investigaci\'on (PID2019-105544GB-I00, PID2022-140510NB-I00 and
RYC2019-027017-I), Xunta de Galicia (CIGUS Network of Research Centers,
Consolidaci\'on 2021 GRC GI-2033, ED431C-2021/22 and ED431F-2022/15),
Junta de Andaluc\'\i{}a (SOMM17/6104/UGR and P18-FR-4314), and the European
Union (Marie Sklodowska-Curie 101065027 and ERDF); USA -- Department of
Energy, Contracts No.~DE-AC02-07CH11359, No.~DE-FR02-04ER41300,
No.~DE-FG02-99ER41107 and No.~DE-SC0011689; National Science Foundation,
Grant No.~0450696, and NSF-2013199; The Grainger Foundation; Marie
Curie-IRSES/EPLANET; European Particle Physics Latin American Network;
and UNESCO.
\end{sloppypar}

\end{acknowledgments}
\twocolumngrid

\appendix

\section{Full results}

We report in \cref{tab:full_results1500} the upper limits obtained with the analysis performed with the SD-1500 array.
In the printed version, only the two objects which show the lowest flux upper limit for each target set are reported.
The machine-readable table available online includes the results for all the 1092 targets considered, in the same format.
Similarly, in \cref{tab:full_results750} the results for the analysis with data coming from the SD-750 array are reported.

\begin{sidewaystable}
    \setlength{\tabcolsep}{6pt}
    
    \centering
 \caption{Upper limits at 95\% confidence level in all the energy bins considered for the analysis with the 1500\;m array. 
 The upper limit for flux is given in km$^{-2}$\,yr$^{-1}$, and the upper limit for energy flux in eV\,cm$^{-2}$\,s$^{-1}$. 
 Each target set is truncated after two entries in the printable version, but the machine-readable table available in the online version of the paper contains all the targets tested.}

 \begin{tabular}{l l l|l l|l l|l l|l l}%
%\tablehead{\colhead{TYPE}&\colhead{RA}&\colhead{Dec}&\colhead{flux\_UL\_1\_2EeV}&\colhead{E\_UL\_1\_2EeV}&\colhead{flux\_UL\_2\_3EeV}&\colhead{E\_UL\_2\_3EeV}&\colhead{flux\_UL\_geq3EeV}&\colhead{E\_UL\_geq3EeV}&\colhead{flux\_UL\_geq1EeV}&\colhead{E\_UL\_geq1EeV}\\}
%\hline%
%TYPE&RA&Dec&flux\_UL\_1\_2EeV&E\_UL\_1\_2EeV&flux\_UL\_2\_3EeV&E\_UL\_2\_3EeV&flux\_UL\_geq3EeV&E\_UL\_geq3EeV&flux\_UL\_geq1EeV&E\_UL\_geq1EeV\\%
\multirow{2}{*}{Class} & \multirow{2}{*}{R.A.}  & \multirow{2}{*}{Dec} & \multicolumn{2}{c|}{$1-2$\,EeV } &  \multicolumn{2}{c}{$2-3$\,EeV}  & \multicolumn{2}{c}{$\geq$ 3\,EeV} & \multicolumn{2}{c}{$\geq$ 1\,EeV}  \\ 
& (deg)	&  (deg)   &  Flux UL & E-flux UL & Flux UL & E-flux UL & Flux UL & E-flux UL &Flux UL & E-flux UL\\
\hline\hline%
CRAB&83.63&22.01&1.75e{-}02&1.28e{-}01&5.20e{-}03&3.80e{-}02&1.18e{-}03&8.63e{-}03&0.014&1.02e{-}01\\%
\hline%
GC&266.42&{-}29.01&4.06e{-}03&2.97e{-}02&1.46e{-}03&1.07e{-}02&8.68e{-}04&6.33e{-}03&0.00365&2.66e{-}02\\%
\hline%
GAMMA\_PSR&168.03&{-}61.13&3.03e{-}03&2.21e{-}02&1.16e{-}03&8.49e{-}03&9.09e{-}04&6.63e{-}03&0.00232&1.69e{-}02\\%
GAMMA\_PSR&270.32&{-}24.85&3.08e{-}03&2.25e{-}02&1.38e{-}03&1.01e{-}02&9.06e{-}04&6.61e{-}03&0.00237&1.73e{-}02\\%
GAMMA\_PSR&... & & & & & & &\\
\hline%
HMXB&107.4&{-}16.1&4.13e{-}03&3.01e{-}02&1.23e{-}03&8.99e{-}03&8.26e{-}04&6.03e{-}03&0.000174&1.27e{-}03\\%
HMXB&266.19&{-}27.23&3.01e{-}03&2.20e{-}02&1.35e{-}03&9.87e{-}03&8.80e{-}04&6.43e{-}03&0.00231&1.69e{-}02\\%
HMXB&...& & & & & & &\\
\hline%
LMXB&270.29&{-}25.08&2.91e{-}03&2.13e{-}02&1.52e{-}03&1.11e{-}02&9.04e{-}04&6.60e{-}03&0.0022&1.61e{-}02\\%
LMXB&290.65&{-}17.28&3.33e{-}03&2.43e{-}02&1.36e{-}03&9.93e{-}03&6.55e{-}04&4.78e{-}03&0.00239&1.75e{-}02\\%
LMXB&...& & & & & & &\\
\hline%
MGN&257.2&{-}40.15&5.90e{-}03&4.30e{-}02&6.09e{-}04&4.44e{-}03&7.95e{-}04&5.80e{-}03&0.00349&2.55e{-}02\\%
MGN&251.79&{-}45.87&5.54e{-}03&4.04e{-}02&9.39e{-}04&6.85e{-}03&7.64e{-}04&5.58e{-}03&0.00363&2.65e{-}02\\%
MGN&...& & & & & & &\\
\hline%
MQSR&269.5&{-}25.13&3.24e{-}03&2.36e{-}02&1.52e{-}03&1.11e{-}02&1.05e{-}03&7.70e{-}03&0.00299&2.18e{-}02\\%
MQSR&237.5&{-}56.07&3.78e{-}03&2.76e{-}02&1.22e{-}03&8.89e{-}03&1.19e{-}03&8.72e{-}03&0.00354&2.58e{-}02\\%
MQSR&...& & & & & & &\\
\hline%
PeV&273.52&{-}17.31&5.44e{-}03&3.97e{-}02&1.51e{-}03&1.10e{-}02&4.95e{-}04&3.62e{-}03&0.00342&2.50e{-}02\\%
PeV&273.72&{-}16.62&6.91e{-}03&5.04e{-}02&9.14e{-}04&6.67e{-}03&6.62e{-}04&4.84e{-}03&0.00397&2.90e{-}02\\%
PeV&...& & & & & & &\\
\hline%
TeV\_OTHER&251.71&{-}45.82&5.26e{-}03&3.84e{-}02&9.40e{-}04&6.86e{-}03&7.65e{-}04&5.58e{-}03&0.0035&2.55e{-}02\\%
TeV\_OTHER&278.21&{-}9.38&3.34e{-}03&2.44e{-}02&2.70e{-}03&1.97e{-}02&1.28e{-}03&9.31e{-}03&0.00364&2.65e{-}02\\%
TeV\_OTHER&...& & & & & & &\\
\hline%
TeV\_PWN&84.43&{-}69.17&3.16e{-}03&2.31e{-}02&1.51e{-}03&1.10e{-}02&8.05e{-}04&5.87e{-}03&0.00269&1.96e{-}02\\%
TeV\_PWN&215.04&{-}60.76&2.91e{-}03&2.12e{-}02&1.49e{-}03&1.08e{-}02&1.37e{-}03&1.00e{-}02&0.00342&2.50e{-}02\\%
TeV\_PWN&...& & & & & & &\\
\hline%
TeV\_UNID&257.1&{-}41.09&3.43e{-}03&2.50e{-}02&1.09e{-}03&7.94e{-}03&9.22e{-}04&6.73e{-}03&0.00305&2.23e{-}02\\%
TeV\_UNID&277.24&{-}9.99&2.33e{-}03&1.70e{-}02&2.84e{-}03&2.08e{-}02&9.06e{-}04&6.62e{-}03&0.00341&2.49e{-}02\\%
TeV\_UNID&...& & & & & & &\\
\hline%
\end{tabular}%
    \label{tab:full_results1500}
\end{sidewaystable}

\begin{sidewaystable}
    \setlength{\tabcolsep}{6pt}
    
    \centering
    \caption{Upper limits at 95\% confidence level in all the energy bins considered for the analysis with the 750\;m array.
    The upper limit for flux is given in km$^{-2}$\,yr$^{-1}$, and the upper limit for energy flux in eV\,cm$^{-2}$\,s$^{-1}$.
    Each target set is truncated after two entries in the printable version, but the machine-readable table available in the online version of the paper contains all the targets tested.}

 \begin{longtable}{l l l|l l|l l|l l|l l}%
%\tablehead{\colhead{TYPE}&\colhead{RA}&\colhead{Dec}&\colhead{flux\_UL\_1\_2EeV}&\colhead{E\_UL\_1\_2EeV}&\colhead{flux\_UL\_2\_3EeV}&\colhead{E\_UL\_2\_3EeV}&\colhead{flux\_UL\_geq3EeV}&\colhead{E\_UL\_geq3EeV}&\colhead{flux\_UL\_geq1EeV}&\colhead{E\_UL\_geq1EeV}\\}
%\hline%
%TYPE&RA&Dec&flux\_UL\_1\_2EeV&E\_UL\_1\_2EeV&flux\_UL\_2\_3EeV&E\_UL\_2\_3EeV&flux\_UL\_geq3EeV&E\_UL\_geq3EeV&flux\_UL\_geq1EeV&E\_UL\_geq1EeV\\%
\multirow{2}{*}{Class} & \multirow{2}{*}{R.A.}  & \multirow{2}{*}{Dec} & \multicolumn{2}{c|}{$0.1-0.2$\,EeV } &  \multicolumn{2}{c}{$0.2-0.3$\,EeV}  & \multicolumn{2}{c}{$\geq 0.3\,$EeV} & \multicolumn{2}{c}{$\geq 0.1\,$EeV}  \\ 
& (deg)	& (deg)    &  Flux UL & E-flux UL & Flux UL & E-flux UL & Flux UL & E-flux UL &Flux UL & E-flux UL\\
\hline\hline%
\hline%
GAMMA\_PSR&153.05&{-}42.59&2.00e{-}01&1.46e+00&2.79e{-}01&2.04e+00&4.60e{-}02&3.36e{-}01&0.113&8.27e{-}01\\%
GAMMA\_PSR&263.15&{-}31.52&3.41e{-}01&2.49e+00&6.67e{-}02&4.87e{-}01&1.79e{-}01&1.31e+00&0.323&2.35e+00\\%
GAMMA\_PSR&...& & & & & & &\\
\hline%
HMXB&104.57&{-}7.21&1.40e+00&1.02e+01&2.36e{-}01&1.72e+00&6.95e{-}02&5.08e{-}01&0.92&6.72e+00\\%
HMXB&190.71&{-}63.06&6.33e{-}01&4.62e+00&2.89e{-}01&2.11e+00&2.68e{-}01&1.96e+00&1.08&7.90e+00\\%
\hline%
LMXB&267.72&{-}31.09&4.92e{-}01&3.59e+00&1.18e{-}01&8.62e{-}01&1.48e{-}01&1.08e+00&0.425&3.10e+00\\%
LMXB&260.85&{-}28.63&4.60e{-}01&3.35e+00&1.37e{-}01&1.00e+00&1.50e{-}01&1.10e+00&0.435&3.17e+00\\%
LMXB&...& & & & & & &\\
\hline%
MSEC&211.34&{-}46.93&5.37e{-}01&3.92e+00&1.99e{-}01&1.46e+00&7.55e{-}02&5.51e{-}01&0.431&3.14e+00\\%
MSEC&260.85&{-}28.63&4.60e{-}01&3.35e+00&1.37e{-}01&1.00e+00&1.50e{-}01&1.10e+00&0.435&3.17e+00\\%
MSEC&...& & & & & & &\\
\hline%
PeV&98.48&17.77&6.94e+00&5.07e+01&1.08e+00&7.85e+00&2.99e{-}01&2.19e+00&4.83&3.53e+01\\%
\hline%
TeV\_OTHER&128.84&{-}45.18&4.96e{-}01&3.62e+00&1.23e{-}01&8.98e{-}01&1.22e{-}01&8.88e{-}01&0.45&3.29e+00\\%
TeV\_OTHER&133.0&{-}46.37&6.35e{-}01&4.63e+00&2.90e{-}01&2.12e+00&9.05e{-}02&6.60e{-}01&0.636&4.64e+00\\%
TeV\_OTHER&...& & & & & & &\\
\hline%
TeV\_PWN&128.75&{-}45.6&4.95e{-}01&3.62e+00&1.23e{-}01&8.99e{-}01&1.83e{-}01&1.33e+00&0.506&3.69e+00\\%
TeV\_PWN&98.12&17.37&5.69e+00&4.15e+01&1.07e+00&7.82e+00&1.83e{-}01&1.34e+00&3.85&2.81e+01\\%
TeV\_PWN&...& & & & & & &\\

\hline%
\end{longtable}%
    \label{tab:full_results750}

\end{sidewaystable}

\bibliography{refs}{}
\bibliographystyle{aasjournalv7}

\end{document}